\documentclass[twocolumn]{aastex7}
\usepackage{amsmath}
\usepackage{CJK}

\graphicspath{{./}{figures/}}

\begin{document}
\begin{CJK*}{UTF8}{gbsn}

\title{Impact of Cold Jupiter Scattering on the Mean-Motion Resonance of Inner Small Planets}

\author[orcid=0000-0001-6870-3114,gname=Kangrou, sname='Guo']{Kangrou Guo (郭康柔)}
\affiliation{Tsung-Dao Lee Institute, Shanghai Jiao Tong University, 1 Lisuo Road, Shanghai 201210, China}
\email[show]{carol.guo@sjtu.edu.cn}  

\author[orcid=0000-0002-2546-2012,gname=Xiumin, sname='Huang']{Xiumin Huang (黄秀敏)}
\affiliation{Tsung-Dao Lee Institute, Shanghai Jiao Tong University, 1 Lisuo Road, Shanghai 201210, China}
\email[]{xm_huang@sjtu.edu.cn}

\author[orcid=0000-0002-1934-6250,gname=Dong, sname='Lai']{Dong Lai (赖东)} 
\affiliation{Tsung-Dao Lee Institute, Shanghai Jiao Tong University, 1 Lisuo Road, Shanghai 201210, China}
\affiliation{Center for Astrophysics and Planetary Science, Department of Astronomy, Cornell University, Ithaca, NY 14853, USA}
\email[show]{donglai@sjtu.edu.cn}

\begin{abstract}

A key feature of close-in, multiple super-Earth (SE) systems is the tendency for adjacent planet pairs to lie just wide of low-order mean-motion resonances (MMR). This period ratio distribution has motivated numerous theoretical studies, particularly those invoking post-disk processes that perturb initially resonant architectures.
We investigate whether orbital instability among cold Jupiters (CJs) can perturb inner SE systems initially in MMR. We show that a single pericenter passage of a highly eccentric CJ can disrupt inner resonances once a critical perturbation strength is exceeded, increasing the libration amplitude of the resonant angles. However, N-body simulations show that deep penetration of CJs into the inner system is uncommon, with $\lesssim$10-20\% of cases reaching $\lesssim10\%$ of the initial semi-major axis of the innermost CJ.
Motivated by these results, we use secular perturbation theory to quantify the impact of time-dependent forcing from scattering CJs on the eccentricity and resonant-angle evolution of inner SEs. We find that for typical systems (e.g., with SEs at $\sim0.1$ au and CJs at a few au), such forcing can efficiently disrupt resonances, driving resonance-angle circulation in most systems ($\gtrsim60\%$ for 2:1 and $\sim85\%$ for 3:2 configurations). Thus, even when the ``final'' CJ has little effect on the ``current'' SEs, its earlier scattering history can leave significant imprints on the system architecture. This mechanism, and similar ones involving more abundant cold Neptunes, provide a natural source of dynamical ``kicks'' and offer a pathway for producing the observed trough–peak structure in the period ratio distribution of Kepler multi-planet systems.

\end{abstract}


\section{Introduction} \label{sec:intro}

The period-ratio distribution of close-in, multiple super-Earth (SE) systems observed by the Kepler spacecraft exhibits a characteristic trough–peak structure near low-order mean-motion resonances (MMR), with a deficit of planet pairs just interior to the resonance and a pile-up just exterior \citep{Lissauer_2011, Fabrycky_2014}. This systematic offset from exact commensurability is most prominent near the 2:1 and 3:2 resonances \citep{Steffen_2015}.
An important question arises: what dynamical processes operating after planet formation can transform initially resonant systems into the near-resonant architectures observed today?

A crucial observational clue comes from age-resolved exoplanet demographics. Using planetary systems with well-constrained stellar ages, \citet{Dai_2024} showed that resonant or near-resonant configurations are significantly more common in young systems than in older ones: the fraction of systems hosting at least one adjacent planet pair near a first- or second-order commensurability decreases from $\sim 86$\% for systems younger than 100 Myr, to $\sim 38$\% for systems aged 100 Myr–1 Gyr, and further to $\sim 23$\% for mature systems older than 1 Gyr. Complementary evidence comes from stellar kinematics: \citet{Hamer_2024} found that host stars of resonant Kepler multi-planet systems exhibit kinematically cooler Galactic velocity distributions than non-resonant systems, indicating that the former have systematically younger ages.
These results strongly suggest that the inner multi-planet systems emerge from protoplanetary disks in compact resonant chains, which are progressively disrupted over $\sim 100$Myr timescale by post-disk dynamical processes.

Other properties of the (near-)resonant and non-resonant systems also support the post-disk instability scenario. Many planet pairs near first-order resonances exhibit non-zero free eccentricities \citep{Lithwick_2012, Wu_Lithwick_2013, Hadden_2014, Hadden_2017}, which are difficult to reconcile with disk-phase mechanisms such as resonant repulsion by eccentricity damping \citep{Lithwick_Wu_2012, Goldreich_Schlichting_2014,Deck_2015}. The presence of free eccentricity therefore points to dynamical excitation after gas dispersal. 
Moreover, planets in or near resonance tend to be ``puffier" compared with those in non-resonant systems \citep{Leleu_2024}, pointing to either an original formation site beyond the snow line or avoidance of the later giant impact stage \citep[e.g.,][]{Chen_2024}. Taken together, these observations indicate that initially resonant chains are subsequently perturbed and displaced from resonance, producing the observed trough–peak structure in the period-ratio distribution.

Motivated by these observations, a number of theoretical studies have explored how giant impacts and late-stage instabilities can break resonant chains. Numerical experiments show that compact resonant configurations assembled during disk migration, when properly ``kicked'' after gas removal, can become unstable, leading to planet–planet scattering, and the resulting planet collision or ejection can shift the system away from exact commensurability \citep[e.g.,][]{Izidoro_2017,Izidoro_2021,Goldberg_2022,Ghosh_2024,Liveoak_2024,Li_rixin_2025}. These studies collectively establish that post-disk instabilities, when triggered, are capable of erasing resonant chains and reproducing the observed near-resonant architectures.

However, resonant chains formed during disk migration are often long-term stable in isolation \citep[e.g.,][]{Izidoro_2017, Tamayo_2017, Lammers_2024}. 
An additional perturbation (a dynamical ``kick”) is required to initiate resonance breaking and subsequent giant impacts. A wide range of such triggers has been proposed, including planet atmospheric mass loss due to photoevaporation or stellar mass loss \citep{Matsumoto_2020},
scattering of SEs by in-situ planetesimals \citep{Wu_2024, Hadden_2026}, 
planetesimal flyby \citep{Li_jiaru_2025_arxiv}, excitation by outer eccentric embryos \citep{Ogihara_2026}, destabilization by unseen outer remnant chains \citep{Goldberg_2025}, and secular forcing from distant giant planets \citep[e.g.,][]{Rodet_2021, Choksi_2023}.

Among these possibilities, perturbations from distant giant planets, commonly referred to as cold Jupiters (CJs), are particularly intriguing. 
About 30\% of SE systems host external CJ companions \citep{Zhu_2018}.
Radial velocity (RV) surveys indicate that CJs possess a broad eccentricity distribution with a moderate mean eccentricity of $\sim 0.23$ \citep[e.g.,][]{Winn_2015,Rosenthal_2024,Kane_2024}.
These observations are consistent with a history of orbital instability and planet–planet scattering \citep[e.g.,][]{Juric_2007, Chatterjee_2008}. Previous studies have shown, however, that the secular perturbation from a single, dynamically settled giant planet is generally too weak to significantly disrupt inner MMRs, unless the giant planet lies unusually close to the inner system \citep{Choksi_2023}. Given that typical super-Earth (SE) systems reside at $\lesssim 0.3$ au while CJs are commonly located beyond the snow line, such large orbital separations tend to suppress direct secular forcing.
Crucially, the observed CJ population may represent only the survivors of a more violent scattering history. In systems that initially hosted multiple giant planets, orbital instability can lead to the ejection of one or more planets. During the ejection process, the escaping planet undergoes a prolonged phase of eccentricity variation and strongly time-dependent apsidal motion, and can also wander close to the inner SE system. The resulting secular forcing can be substantially stronger and more stochastic than that produced by a single, stable giant planet \citep{Pu_2021}. Such perturbations have the potential to excite eccentricities in the inner system, alter the amplitudes of resonant angles, and ultimately destabilize resonant chains.

In this work, we explore the impact of CJ scattering histories on the stability of inner mean-motion resonances. By explicitly modeling the perturbations exerted on the inner system during giant planet ejection events, we assess whether this mechanism can provide a natural and efficient pathway for disrupting the resonant configuration, potentially responsible for triggering resonant-chain instability and producing the near-resonant architectures observed in close-in SE systems.

This paper is organized as follows. Section \ref{sec:part_I} quantifies how a single close encounter—arising from the pericenter passage of a highly eccentric CJ—modifies the resonant angles and period ratio of an inner SE pair initially in 2:1 or 3:2 MMR. Section \ref{sec:scatter} presents results from our scattering experiments, focusing on the probability that an unstable CJ penetrates the inner system. Section \ref{sec:secular} then examines how the long-term secular forcing from two mutually scattering CJs influences the inner MMR configurations. In Section \ref{sec:conclusion}, we summarize our findings and discuss potential caveats and broader implications of these findings.

\section{Critical distance to destroy the resonance} \label{sec:part_I}

If a giant planet is scattered onto a highly eccentric orbit through planet–planet interactions, its pericenter can become small enough for the planet to penetrate the inner system during its radial excursion. When the pericenter is sufficiently close, the giant planet can significantly perturb the inner system through a single close encounter. Here we quantify the strength of such a “kick” as a function of the separation between the inner SE pair and the pericenter distance of the incoming giant. We focus on the effect of a single pericenter passage of a high-eccentricity giant planet on an inner SE pair locked in first-order MMRs (including 2:1 and 3:2), varying the giant planet’s pericenter distance to determine how close it must approach to induce substantial disturbance to the resonance. We examine two quantities central to resonant dynamics. The first one is the resonant angle: 
\begin{equation}
    \phi_{1,2} = (q+1)\lambda_2 - q\lambda_1 - \varpi_{1,2},
    \label{eq:phi12}
\end{equation}
where $\lambda_1,\lambda_2$ are the mean longitudes and $\varpi_1, \varpi_2$ are the longitudes of pericenter of the two planets. 
The other is the fractional deviation from the $(q+1):q$ commensurability, 
\begin{equation}
    \Delta = \frac{q}{q+1}\frac{P_2}{P_1} - 1,
    \label{eq:Delta}
\end{equation}
where $P_1$ and $P_2$ are the orbital periods of the two planets.

For simplicity, we restrict the models in Section \ref{sec:part_I} to coplanar, including the SEs and the giant planet.
We first let the two SEs undergo convergent migration (phase I) and enter MMR using the \texttt{modify\_orbits\_forces} module in \texttt{REBOUNDx} \citep{Tamayo_2020_rbx}. Both SEs have a planet–to–star mass ratio of $\mu_1 = \mu_2 \simeq 2.25 \times 10^{-5}$ ($7.5~M_\oplus$ around a Solar-mass star), with the outer SE starting at a semi-major axis of $a_2 = 0.2$ au. The planets are initially placed slightly wide of resonance by 3\% and have very small eccentricities of $10^{-4}$. During convergent migration, the outer SE experiences semi-major axis damping with a timescale $\tau_a$, while both planets experience eccentricity damping with a timescale $\tau_e$. The equilibrium eccentricity reached in resonance is $e \sim \sqrt{\tau_e/\tau_a}$. Following \citet{Choksi_2023}, we use equations (49)–(51) in \citet{Terquem_2019} to determine the ratio $\tau_e/\tau_a$ that yields an equilibrium $\Delta$ of approximately 1\%.

Phase I is integrated for $10^6/n_{1,0}$, where $n_{1,0}$ is the initial orbital frequency of the inner SE. We then gradually remove the damping forces over a timescale of $5\tau_e$. After confirming that the SE pair has settled into resonance (with $\Delta \simeq 1\%$), we continue integrating the system for an additional $1000 P_{1,0}$, where $P_{1,0}$ is the initial orbital period of the inner SE.
At the end of phase I, $\phi_1$ and $\phi_2$ typically librate around 0 and $\pi$, respectively, with very small libration amplitude.
For SEs in 2:1 MMR, the libration amplitudes of $\phi_1$ and $\phi_2$ are $\simeq 0.02$ and $\simeq 0.20$ radian. For the 3:2 MMR case, the libration amplitudes are $\simeq 0.12$ radian for both angles.

In phase II, we introduce an outer CJ with planet–to–star mass ratio $\mu_3$ on a highly eccentric orbit. Starting from its apocenter, the CJ approaches the system, passes through pericenter once, and the integration is halted upon its return to apocenter (one CJ orbital period). The change in the deviation from commensurability, $\Delta$, is measured as the difference between the values of $\Delta$ at the beginning and end of phase II. The change in the libration amplitudes of the resonant angles, $\Delta\phi_{1,2}$, is measured as the difference between the maximum amplitude of $\phi_{1,2}$ after the pericenter passage and the amplitude at the beginning of phase II (at which point the libration amplitudes are very small due to prior convergent migration into resonance). 

We define a dimensionless parameter, the perturbing acceleration ratio, as  
\begin{equation}
  \eta \equiv \frac{Gm_3/d^2}{GM_*/a_2^2} = \mu_3 \left(\frac{a_2}{d}\right)^2,
  \label{eq:eta}
\end{equation}
where $d=r_{\rm peri}-a_2$ is the separation between the CJ's pericenter and the outer SE. This parameter measures how strongly a close approach by the CJ perturbs the outer SE away from its Keplerian orbit. 

We explore eight values of $\eta$, sampled uniformly in logarithmic space from $10^{-4}$ to 0.3. The CJ’s mass is fixed at $\mu_3 = 0.001$. For this mass, the chosen range of $\eta$ corresponds to pericenter distances between roughly 0.21 and 0.83 au, assuming a CJ semi-major axis of 3 au. For each $\eta$, we run 50 simulations with the initial orbital phases of the inner SEs drawn randomly from $[0, 2\pi]$. The full set of simulation parameters is summarized in Table \ref{tab:param}.

\begin{deluxetable*}{cccc}
\tablewidth{0pt}
\tablecaption{Parameters of simulations} \label{tab:param}
\tablehead{
\colhead{batch} & \colhead{$e_3$} & \colhead{$r_{\rm{peri}}$ (au)} & \colhead{$\eta$}
}
\startdata
1 & 0.7225 & 0.8325 & 0.0001\\
2 & 0.8116 & 0.5651 & 0.0003\\
3 & 0.8667 & 0.4000 & 0.001\\
4 & 0.8948 & 0.3155 & 0.003\\
5 & 0.9123 & 0.2631 & 0.01\\
6 & 0.9212 & 0.2365 & 0.03\\
7 & 0.9267 & 0.2200 & 0.1\\
8 & 0.9295 & 0.2115 & 0.3\\
\enddata
\tablecomments{The eccentricity $e_3$, pericenter distance $r_{\rm{peri}}$ of the incoming CJ and the corresponding value of the perturbing acceleration ratio $\eta$ (see Eq. (\ref{eq:eta}) in batch of simulations (50 runs in each batch).}
\end{deluxetable*}

\subsection{2:1 MMR case}

\begin{figure*}
    \centering
    \includegraphics[width=\textwidth]{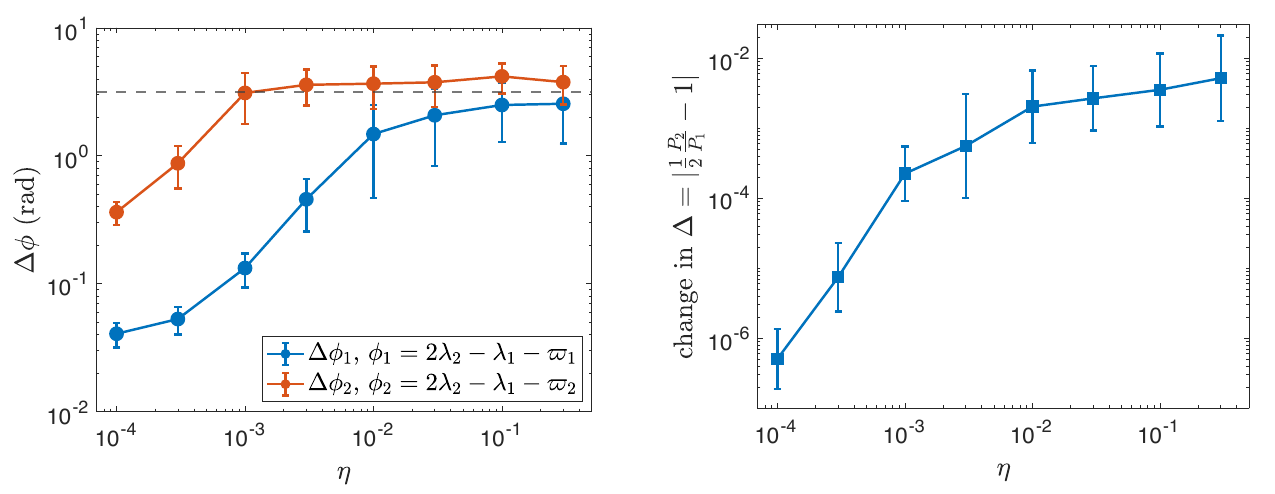}
    \caption{Impact of an incoming giant planet on the inner 2:1 MMR pair. \textit{Left:} change in the oscillation amplitudes of the resonant angles $\phi_1 = 2\lambda_2-\lambda_1-\varpi_1$ (blue) and $\phi_2 = 2\lambda_2-\lambda_1-\varpi_2$ (red) after the pericenter passage of the incoming giant planet. Each point shows the mean value from 50 simulations for a given value of $\eta$ (see Eq. (\ref{eq:eta})), the dimensionless perturbing acceleration ratio. Error bars show the standard deviation. The dashed gray line in the left panel highlights where $\Delta \phi = \pi$, i.e., the transition from libration to circulation. \textit{Right:} change in the deviation from exact 2:1 orbital period commensurability following the pericenter passage.}
    \label{fig:I-mmr21}
\end{figure*}

Figure \ref{fig:I-mmr21} shows the impact of a single pericenter passage of a high-eccentricity giant planet as a function of $\eta$. The left panel displays the change in the libration amplitude $\Delta\phi_{1,2}$ of the resonant angles of the inner SE pair. When $\Delta\phi$ exceeds $\pi$, the resonant angle transitions from libration to circulation. The right panel shows the corresponding change in $\Delta$. Each point represents the mean value of 50 simulations, with error bars showing the standard deviation.

A few trends can be identified from the left panel. First, both resonant angles exhibit larger oscillation amplitudes as $\eta$ increases. Second, $\Delta\phi_2$ is systematically larger than $\Delta\phi_1$. This is expected because the outer SE lies closer to the pericenter of the incoming CJ and is therefore more strongly affected by the gravitational kick. When $\eta \gtrsim 10^{-3}$, $\Delta\phi_2$ reaches $\pi$, indicating circulation of $\phi_2$, whereas $\phi_1$ remains bounded below 1 rad across the parameter range explored here.

The right panel further shows that the deviation from 2:1 commensurability increases monotonically with $\eta$. However, the overall magnitude of the shift remains small: after resonance capture, $\Delta$ is of order 0.01, while the maximum change in $\Delta$ induced by the pericenter passage is only $\simeq 0.005$ at $\eta = 0.3$. Therefore, although a close passage of an incoming CJ can significantly excite the resonant angles (even driving one of them into circulation), it does not substantially alter the orbital period ratio of the inner SE pair.

\subsection{3:2 MMR case}

\begin{figure*}
    \centering
    \includegraphics[width=\textwidth]{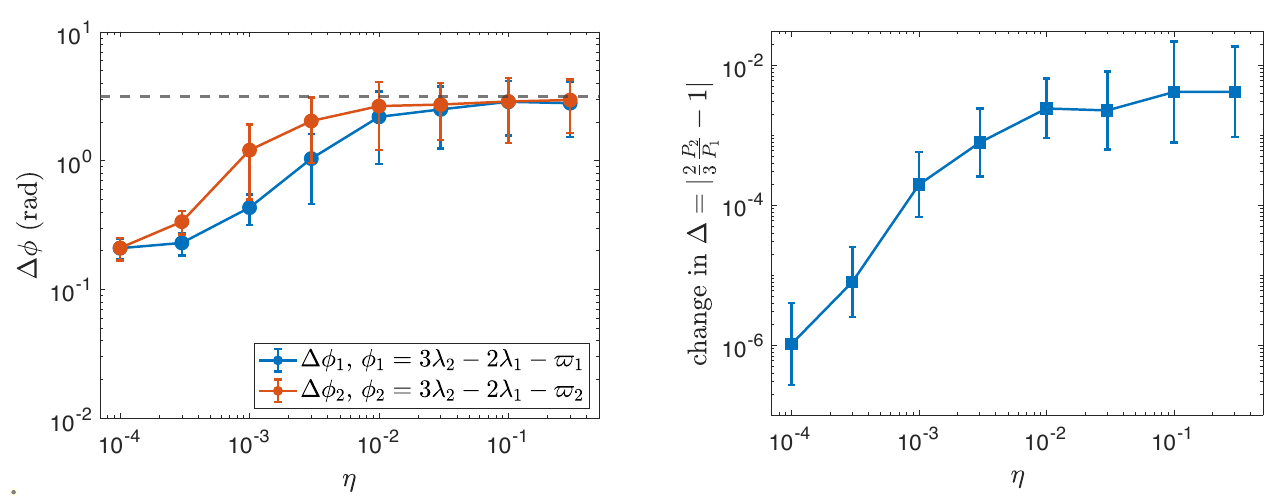}
    \caption{Same as Figure \ref{fig:I-mmr21}, but for the case where the inner SEs are in a 3:2 MMR configuration.}
    \label{fig:I-mmr32}
\end{figure*}

We also examine the effect of an incoming CJ on the SE pair when they are in a 3:2 MMR. Figure \ref{fig:I-mmr32} shows the results. In the left panel, the rise of $\Delta\phi$ with increasing $\eta$ mirrors the behavior seen in the 2:1 case (Figure \ref{fig:I-mmr21}). However, the magnitudes differ: $\Delta\phi_1$ is larger by a factor of a few compared to the 2:1 case, while $\Delta\phi_2$ is smaller by a factor of 1–2. Although $a_2$ is fixed at 0.2 au in both cases, the SEs occupy more compact orbits in the 3:2 resonance. As a result, the resonance is “tighter,” with stronger mutual interactions that cause the perturbation from the CJ’s passage to be distributed more evenly between the two planets. This sharing of the perturbation increases $\Delta\phi_1$ while reducing $\Delta\phi_2$ relative to the 2:1 case.

The right panel shows the fractional deviation from exact 3:2 commensurability. Both the trend and the absolute magnitude of the deviation are comparable to those in the 2:1 case, largely because $a_2$ remains fixed.

Overall, across the parameter ranges explored, a single close passage of an incoming CJ can substantially disturb the resonant angles of an inner SE pair in MMR, but it has only a minor impact on their period ratio.

\section{Scattering simulations of giant planets} \label{sec:scatter}

Knowing how a single close encounter at the CJ’s pericenter passage perturbs an inner SE pair in resonance, the next question is: how likely is it for a CJ to be scattered inward, and how close can it approach the inner system? In this section, we address these questions using REBOUND simulations of planet–planet scattering.

\subsection{Scale-free case} \label{subsec:scale-free}

\begin{figure}
    \centering
    \includegraphics[width=0.45\textwidth]{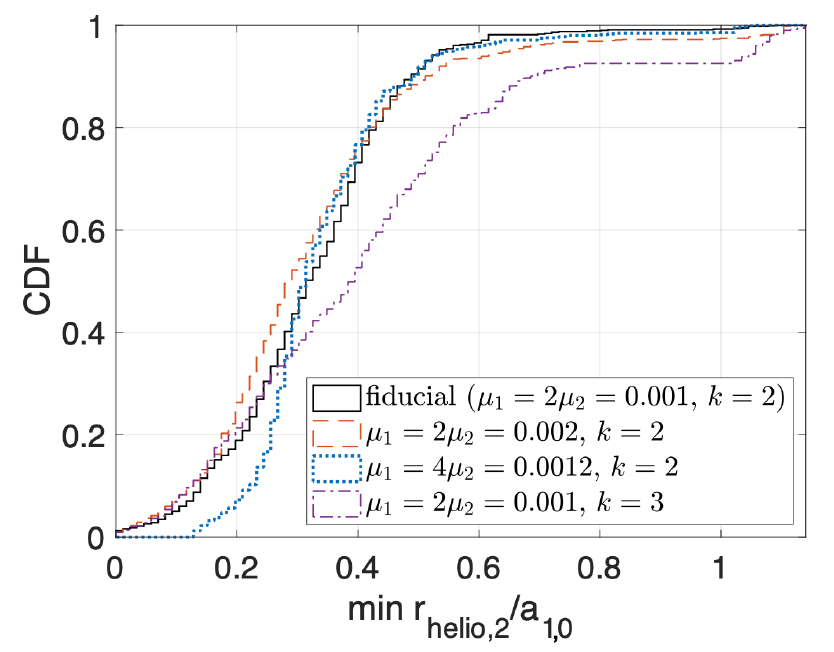}
    \caption{Cumulative distribution of the minimum heliocentric distance of GP2 in four different settings, varying the total mass $\mu_1+\mu_2$ (red dashed histogram), the GP mass ratio $\mu_1/\mu_2$ (blue dotted histogram), and the initial separation factor $k$ (purple dash-dot histogram). The fiducial case is shown by the black solid histogram.}
    \label{fig:scale_free_results}
\end{figure}

For simplicity, we first consider the “scale-free’’ case, in which the two planets are treated as point masses and physical collisions are ignored. The planets are placed on initially unstable, coplanar orbits. The inner giant planet (GP1) is located at $a_{1,0} = 4.28$ au, and the outer giant planet (GP2) is located at $a_{2,0} = a_{1,0} + k r_H$, where $r_H$ is their mutual Hill radius. 
Both planets have a small initial eccentricity of 0.001.
To explore how the initial configuration influences the outcome, we vary the initial separation factor $k$, the mass ratio of the two planets, and their total mass. The fiducial case is set with a planet-to-star mass ratio of $\mu_1=2\mu_2 =0.001$ and an initial separation of $k=2$. 

We focus on the minimum heliocentric distance reached by the outer planet (GP2, the less massive one), because in most cases GP2 is the planet that becomes unbound.

We use the \texttt{REBOUND} code \citep{Rein_2012} with the IAS15 integrator \citep{Rein_2015_IAS15} and integrate each system up to $10^7 P_{1,0}$, where $P_{1,0}$ is the initial orbital period of the inner giant (GP1) \footnote{Note that $P_{1,0}$ here has a different definition than that of the one appeared in Section \ref{sec:part_I}}. The integration is terminated once one planet becomes unbound from the system (i.e., when the planet reaches a heliocentric distance of 1000 au while having positive orbital energy).
\footnote{About 10\% of the systems were discarded because the relative energy error $|\Delta E/E_0|$ exceeds 1\% due to extremely close encounter.}

Figure \ref{fig:scale_free_results} shows the cumulative distribution function (CDF) of the minimum heliocentric distance of GP2 ($r_{\rm helio,2}$) for four sets of initial conditions. For each set of parameters, 500 simulations are performed. Distances are normalized by the initial semi-major axis of GP1, $\min{r_{\rm helio,2}/a_{10}}$.
The CDF demonstrates that the planet–planet mass ratio primarily regulates the low end of the distribution ($\min{r_{\rm helio,2}/a_{1,0}} \lesssim 0.3$), while the initial separation $k$ affects the higher end ($\min{r_{\rm helio,2}/a_{1,0}}\gtrsim 0.3$).
Figure \ref{fig:scale_free_results} also shows that for the $\mu_1=4\mu_2$ case, the minimum $r_{\rm peri,2}$ is always larger than $0.12~a_{1,0}$. This is consistent with the finding in \citet{Rodet_2024}, and can be understood using the conservation of energy and angular momentum.

The probability of reaching $\min(r_{\rm helio,2}/a_{1,0}) \lesssim 0.1$ is less than about 6\% across all simulation sets. This indicates that deep intrusion of CJs into the inner system through scattering is very rare.

\subsection{More realistic scenarios} \label{subsec:3D_scatter}

\begin{figure*}
    \centering
    \includegraphics[width=\textwidth]{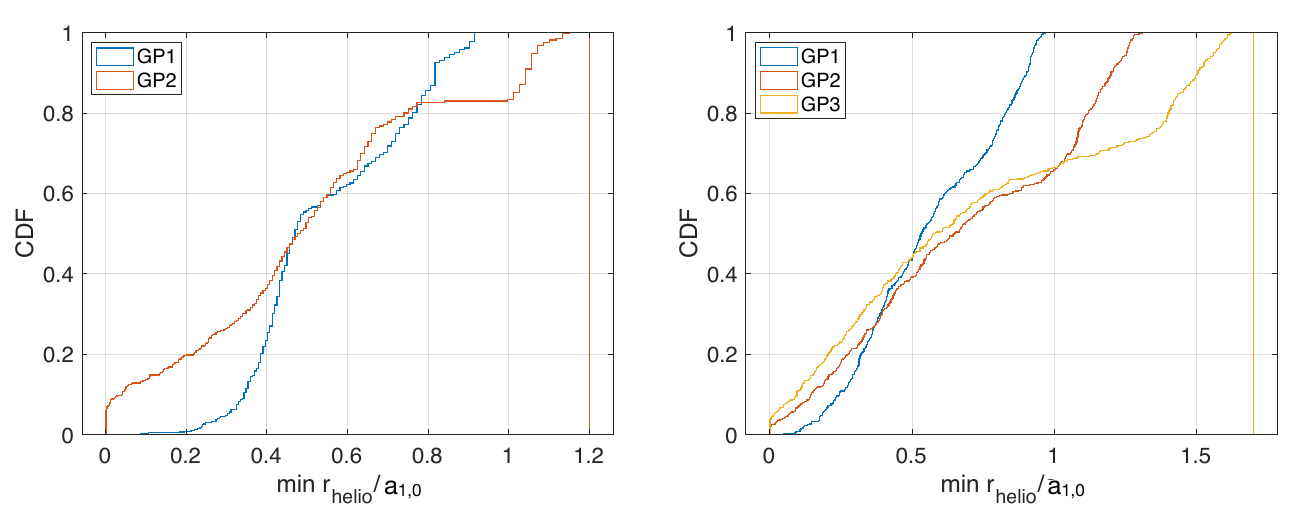}
    \caption{\textit{Left}: Results of 500 two–giant-planet scattering simulations. The curves show the cumulative distribution of the minimum heliocentric distance reached by the inner (blue, GP1) and outer (red, GP2) giant planet. \textit{Right}: Results of 500 three–giant-planet scattering simulations. The blue, red, and yellow curves show the cumulative distribution of the minimum heliocentric distance reached by the innermost (GP1), middle (GP2), and outermost (GP3) planet, respectively.}
    \label{fig:3D_scatter}
\end{figure*}

In this section, we present results from planet–planet scattering simulations under more realistic conditions, where physical collisions are allowed and treated as perfect mergers and the two CJs are initially on slightly misaligned orbits. During each merger event (when the separation of two bodies is less than the sum of their radii), the total mass, linear momentum, and center-of-mass position are conserved.
Tidal effects during close encounters, which increase the effective collision radius of the planet \citep{Li_Jiaru_2021}, are neglected here.

\subsubsection{Two giant planet case}

We first perform 500 scattering simulations with two giant planets. The inner planet GP1 starts at $a_{1,0} = 4.28$ au, and the outer planet GP2 is placed at $a_{2,0} = a_{1,0} + 3r_H$. The planet–to–star mass ratios are $\mu_1 = 0.001 = 2\mu_2$. Both planet have a radius of Jupiter $R_{\rm J}$. Both planets begin with a small inclination $i_0 = 0.1R_H/a_{1,0}$, where $R_H$ is the Hill radius of this planet. The longitude of ascending node, the argument of pericenter, and the initial mean anomaly of the two planets were randomly sampled from 0 to $2\pi$. The integration stops if either a merger or an ejection occurs.
Otherwise the simulations proceed the same way as in Section \ref{subsec:scale-free}.

Among the 500 systems, 55\% result in an ejection of one planet, 43\% experience a merger, and only 2\% remain stable (no merger or ejection before $t_{\rm end}= 3\times 10^7 P_{1,0}$). 
Most collisions occur at $t \sim 10^3 - 10^4$ years, with a long tail extending to $t \sim 10^9$ years, while ejections happen mostly at $t \sim 10^5 - 10^6$ years.

Figure \ref{fig:3D_scatter} (left panel) shows the cumulative distribution of the minimum heliocentric distance reached by both planets. The probability that GP2 approaches within $\min{r_{\rm helio}/a_{1,0}} \simeq 0.1$ is about 14\%, more than twice the probability found in the scale-free case (see Figure \ref{fig:scale_free_results}). 

\subsubsection{Three giant planet case}

Two–planet scattering from nearly circular orbits produces a relatively simple set of outcomes: instabilities typically conclude after a single close encounter, with a narrow branching between one collision and one ejection. The surviving planet rarely exceeds $e \gtrsim 0.8$, so its pericenter does not plunge particularly deep \citep{Ford_2008}.  

In contrast, three–planet scattering exhibits a much wider range of outcomes. Repeated close encounters are common, and these interactions routinely push planets to extremely small pericenter distances and very high eccentricities ($e \gtrsim 0.9$–0.99) before either ejection or long–term survival \citep{Chatterjee_2008, Carrera_2019}.  

To explore these more general configurations, we simulate systems containing three giant planets. The innermost planet GP1 is placed at $a_{1,0}=4.28$ au, and each adjacent pair is initially separated by $3r_H$, where $r_H$ is their mutual Hill radius. The planet–to–star mass ratios are $\mu_1 = 0.002 = 2\mu_2 = 4\mu_3$, and all planets start with a small inclination $i_0 = 0.1R_H/a_{1,0}$. All simulations are integrated until $t_{\rm end} \simeq 10^8$ years.

Across the 500 simulated systems, instability outcomes consist of 38\% collisions and 62\% ejections.  
Most collisions happen at $t \sim 10^3 - 10^4$ years, with a long tail extending to the end of the simulations. Ejections predominantly occur at $t \sim 10^6$ years.

Figure \ref{fig:3D_scatter} (right panel) presents the cumulative distributions of the minimum heliocentric distance reached by each planet. Considering all three planets together, there is roughly a 20\% probability that at least one of them approaches within $\min{r_{\rm helio}/a_{1,0}} \lesssim 0.1$. 
Because collisions are included in the ``realistic'' simulations in Section \ref{subsec:3D_scatter}, no systems exhibit relative energy errors exceeding $10^{-6}$, and no runs are discarded from the analysis.

We emphasize that the scattering experiments presented here are not intended to model fully self-consistent formation scenarios of cold giant planets. Instead, our goal is to characterize the typical dynamical outcome during giant planet scattering events, especially focusing on how close the giant planet(s) can dive into the inner system. The adopted initial conditions are therefore chosen to produce representative post-scattering configurations rather than to reproduce the full formation history. In particular, the initial semi-major axis of 4.28 au is selected such that, following a close scattering event, the surviving giant planet typically settles at $\simeq 3$ au based on energy conservation arguments (e.g., Equation 4 of \citealp{Pu_2021}, hereafter PL21), consistent with the observed peak of cold giant planet occurrence \citep[e.g.,][]{Fernandes_2019, Fulton_2021}. The relatively tight initial spacing (2–3 mutual Hill radii) is likewise adopted to ensure prompt dynamical instability; previous work has shown that the precise initial spacing has limited influence on the statistical properties of scattering outcomes once the onset of instability is reached (PL21). Under this framework, our simulations are designed to probe the range of closest approaches and time-dependent perturbations experienced by inner systems during scattering, rather than to reproduce the detailed dynamical histories of observed cold giant planets.

\section{Secular influence of scattering Cold Jupiters on inner resonance}
\label{sec:secular}

For the majority of systems ($\gtrsim 80$\% in our giant-planet scattering experiments), the CJs do not dive deep into the inner system, and their influence on the inner planets is predominantly secular. In this section, we show how the scattering of two CJs modifies the orbits of an inner SE pair in MMR.

Direct N-body integration of SEs and CJs is computationally expensive because their dynamical timescales differ by several orders of magnitude. Since the perturbations from the CJs on the inner SEs are mainly secular, we instead integrate the secular equations of motion (EoM) for the SEs. We first use REBOUND to run scattering simulations of two CJs (see Section \ref{subsec:scale-free}), extract the time series of their orbital elements at the end of each simulation, and use these as time-dependent secular forcing terms in the EoM of the SEs (listed in Appendix \ref{appendix:EoM}). 
This approach was also adopted in PL21 for SEs not in MMR.
In effect, we apply the evolving apsidal forcing from the CJs to the inner SEs while neglecting the rare deep encounters that violate the secular assumptions. The validity of this approach is discussed in section \ref{subsec:secular_approach}.

\subsection{GP scatter results} \label{subsec:scat_results}

\begin{figure*}
    \centering
    \includegraphics[width=\textwidth]{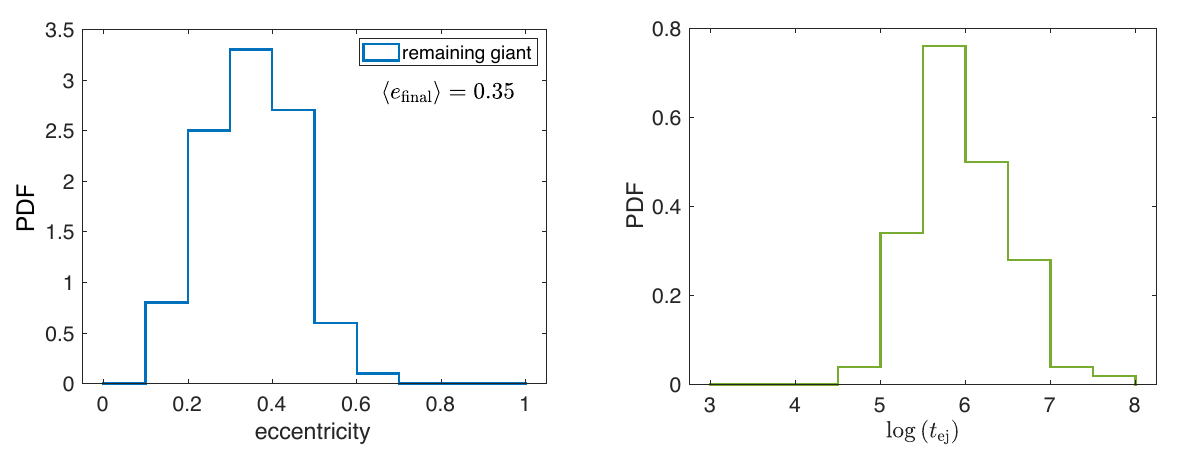}
    \caption{A summary of the giant-planet scattering simulations used for the secular analysis (see Section \ref{sec:secular}). (a) Eccentricity distribution of the surviving giant planet, with a mean value of $\langle e_{\rm final}\rangle = 0.35$. (b) Distribution of the ejection times $t_{\rm ej}$ (in years), shown on a logarithmic scale.}
    \label{fig:scatter_summary_secIII}
\end{figure*}

We focus on how the scattering of the outer giants affects the inner SE pair, and therefore adopt a simplified setup by restricting the simulations to coplanar, no-collision configurations (similar to the scale-free case in Section \ref{subsec:scale-free}). The inner CJ is initialized at $a_3 = 4.28$ au\footnote{In this section, subscripts 3 and 4 refer to the outer CJs, and 1 and 2 to the inner SEs.}, with the two giants separated by $2r_H$. The planet–to–star mass ratios are $\mu_3 = 2\mu_4 = 0.001$, and both planets begin with a small eccentricity of 0.001. We perform 100 REBOUND simulations, each with the initial orbital phases randomly sampled between 0 and $2\pi$. The integrations run up to $t_{\rm end} = 3 \times 10^7 P_{3,0}$, unless one planet is ejected earlier. A planet is considered ejected once its semi-major axis exceeds 1000 au.

The results of these experiments are summarized in Figure \ref{fig:scatter_summary_secIII}. The left panel shows the PDF of the eccentricities of the remaining planet, with the simulations yielding a mean value of $\langle e_{\rm final}\rangle \simeq 0.35$. This is consistent with the empirical scaling from PL21, which predicts $\langle e_{\rm final}\rangle \simeq 0.7\,\mu_4/\mu_3$ (for $\mu_4 \lesssim \mu_3/2$), and a standard deviation of roughly $0.5\,\langle e_{\rm final}\rangle$.

The right panel shows the distribution of the ejection time, $\log t_{\rm ej}$ (in years). Following PL21, we separate out the initial ramp-up phase of eccentricity growth, and define the ejection time $t_{\rm ej}$ as the interval measured from when the apocenter of the inner CJ and pericenter of the outer CJ first lie within one mutual Hill radius, i.e., when $a_4(1-e_4)-a_3(1+e_3)\le r_H$.

Most ejections occur on Myr timescales. As shown by PL21, this ejection timescale depends only weakly on the initial orbital separation once the initial meta-stable ramp-up phase is accounted for.

Out of the 100 simulated systems, 93 maintain a minimum pericenter distance for both planets larger than 0.2 au (i.e., $\min(r_{\rm peri,3,4}) > a_2$), ensuring that the CJ orbits never cross those of the inner SEs. We therefore use the CJ orbital time series from these 93 systems as the secular perturbation input for the EoM of the inner SEs.

\subsection{Impact on resonant pair}

\subsubsection{2:1 MMR case} \label{subsec:2-1mmr}

\begin{figure*}
    \centering
    \includegraphics[width=\textwidth]{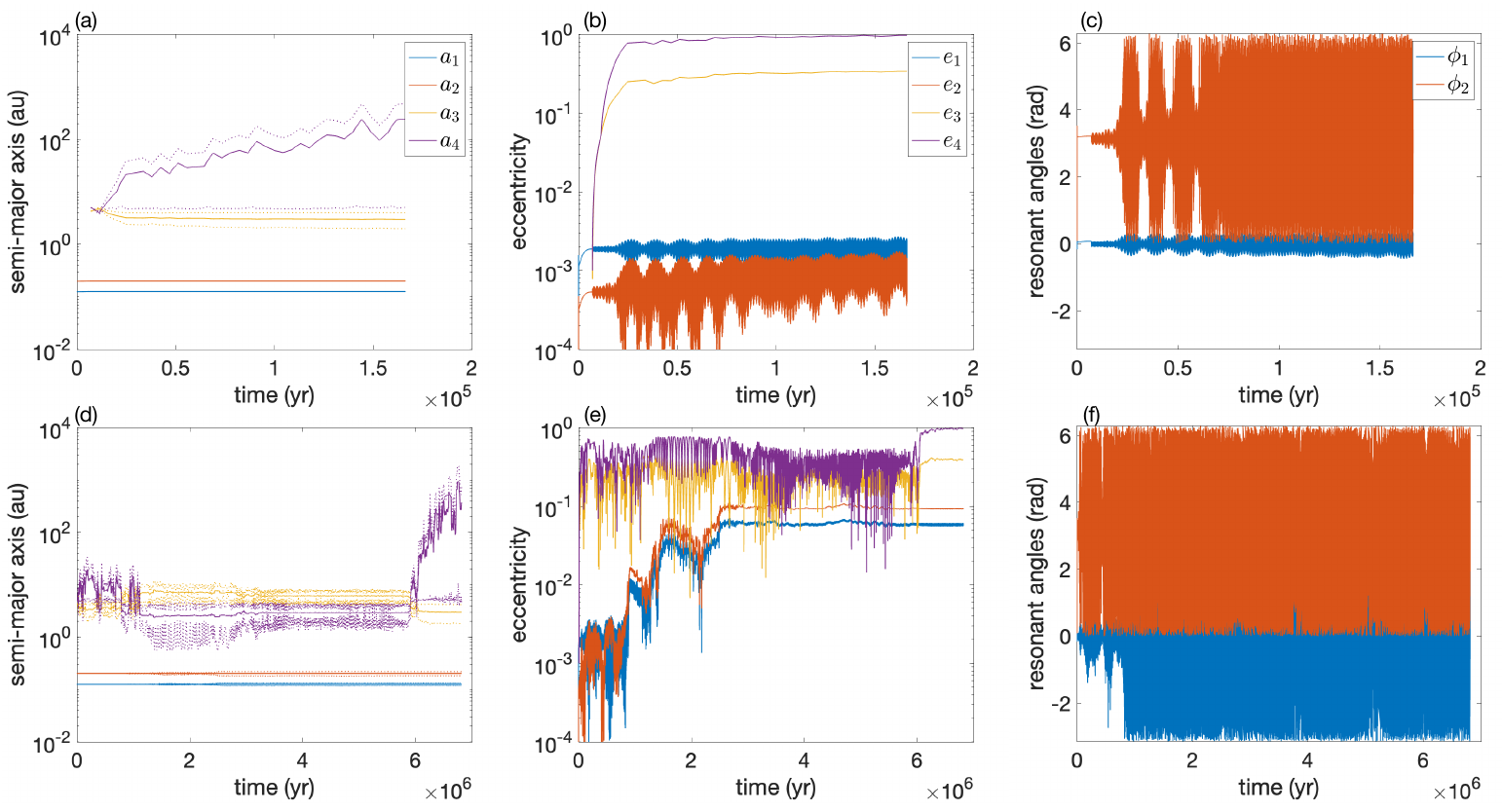}
    \caption{Two examples illustrating how scattering between the outer giant planets affects an inner pair of SEs in a 2:1 MMR. The upper row shows a mildly perturbed case in which the inner SEs’ eccentricities remain largely unexcited. The lower row shows a more violently perturbed case where the inner SEs’ eccentricities increase by more than an order of magnitude. Panels (a)–(c) and (d)–(f) show, respectively, the time evolution of the semi-major axes, the eccentricities, and the resonant angles $\phi_{1,2} = 2\lambda_2 - \lambda_1 - \varpi_{1,2}$. In panels (a) and (d), the dashed lines show the pericenter and apocenter distances.}
    \label{fig:secular_evolution_mmr21}
\end{figure*}

We show two representative systems that undergo mild and violent disruption, corresponding to the upper (panels a–c) and lower rows (panels d–f) of Figure \ref{fig:secular_evolution_mmr21}. In the mildly disrupted case, the two CJs scatter until one of them becomes unbound (purple, P4). Throughout the interaction, the ejected giant (P4) remains exterior to the ejector giant (P3), while its eccentricity gradually approaches unity. The semi-major axes of the inner SEs remain unchanged because we are considering secular effects only. Their eccentricities are barely excited, as the outer giant planets never intrude into the inner system during the scattering. The resonant angle associated with the outer SE ($\phi_2$) transitions to circulation, whereas $\phi_1$ continues to librate with a slightly increased amplitude. This behavior is expected, since the outer resonant angle is more susceptible to perturbations from bodies located exterior to the SEs’ orbits. In this scenario, the resonance is perturbed but remains effectively intact.

In the more violently disrupted case (panels d–f), one of the giant planets (P4) penetrates into the inner system: its semi-major axis becomes smaller than that of the inner giant (P3), and its pericenter drops to within 1 au before it is eventually ejected. This deeper intrusion excites the eccentricities of the inner SEs by nearly two orders of magnitude. Although the SE pair retains stable semi-major axes, both resonant angles transition to circulation once the outer giant (P4) undergoes close orbital exchanges with P3 and enters the inner region. In this case, the resonance is fully broken.

\begin{figure*}
    \centering
    \includegraphics[width=\textwidth]{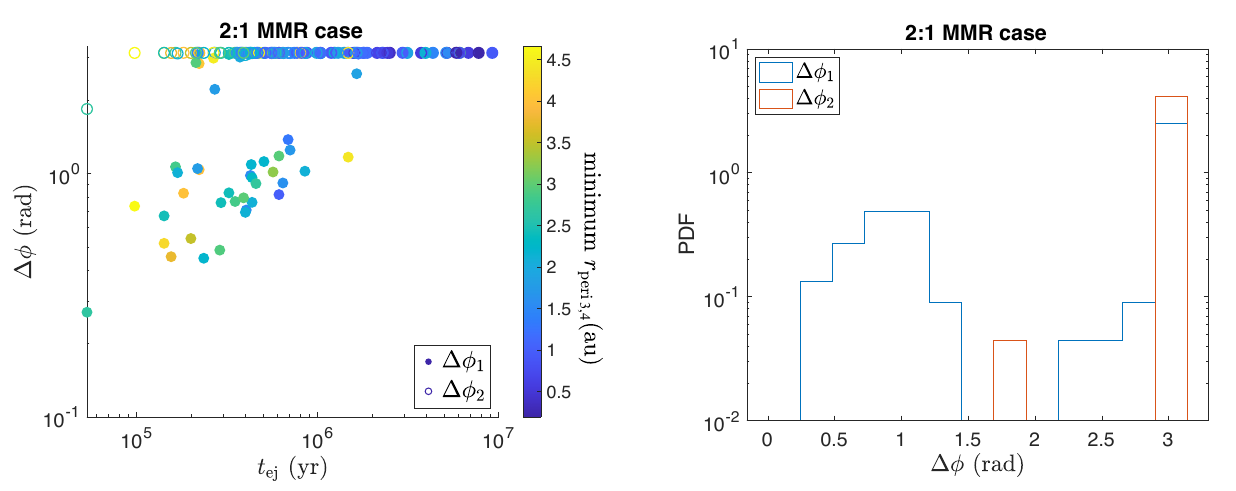}
    \caption{Results of the 93 simulations. \textit{Left}: Scatter plot showing the change in resonant-angle oscillation amplitude, $\Delta\phi$, as a function of the ejection time $t_{\rm ej}$. Filled and open circles denote $\Delta\phi_1$ and $\Delta\phi_2$, respectively. The color scale indicates the minimum pericenter distance reached by the outer giant planets, $r_{\rm peri 3,4}$. \textit{Right}: Distribution of $\Delta\phi_{1,2}$ for all 93 systems.}
    \label{fig:res_angle_mmr21}
\end{figure*}

Out of 93 systems in which the ejected giant planet’s minimum pericenter distance remains larger than the outer super-Earth’s orbit ($r_{\rm peri,2} > a_2$), 65 systems end with $\phi_1$ circulating and 92 systems end with $\phi_2$ circulating. Figure \ref{fig:res_angle_mmr21} summarizes the resulting changes in the resonant angles $\Delta\phi_{1,2}$ (measured within $[0,\pi]$) at the end of each simulation, when one giant planet becomes unbound. 
The left panel shows $\Delta\phi$ as a function of the ejection time $t_{\rm ej}$. Systems with longer ejection times tend to reach smaller $r_{\rm peri 3,4}$, leading to larger $\Delta\phi$. This trend is consistent with Figure \ref{fig:secular_evolution_mmr21}: the longer a giant planet undergoes scattering before ejection, the deeper it intrudes into the inner system. Such intrusion strengthens the secular perturbations on the inner super-Earths, causing stronger eccentricity excitation and enhanced resonant-angle disruption.
The right panel shows the PDF of $\Delta\phi$. In most systems, both resonant angles ultimately circulate. About one-third of the systems retain libration in $\phi_1$, but with moderate shifts ($0 < \Delta\phi_1 < 1.5$). Nearly all systems end with $\phi_2$ circulating.  

\subsubsection{3:2 MMR case} \label{subsec:3-2mmr}

\begin{figure*}
    \centering
    \includegraphics[width=\textwidth]{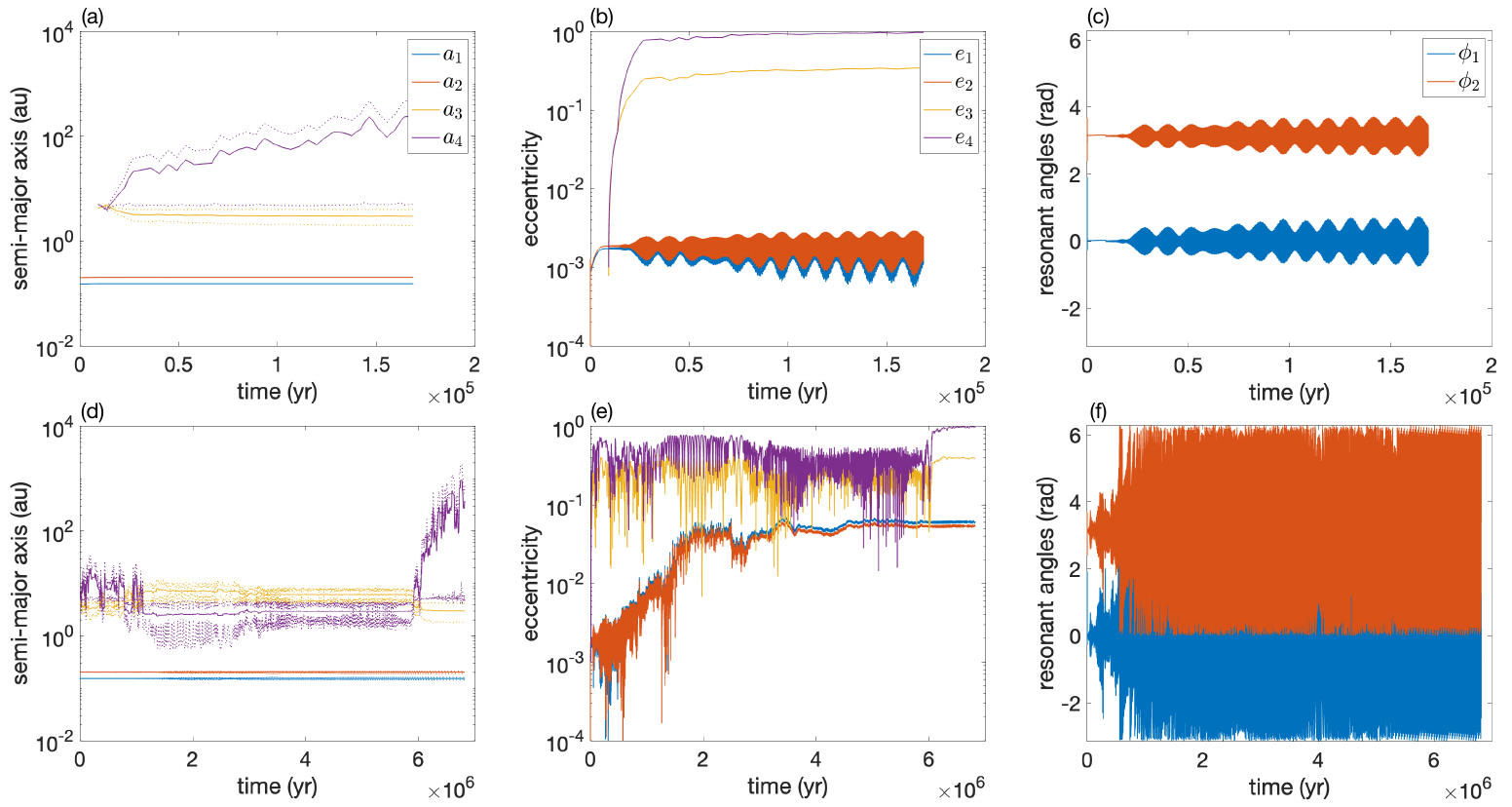}
    \caption{Same as Figure \ref{fig:secular_evolution_mmr21}, but for the case in which the inner SEs begin in a 3:2 MMR.}
    \label{fig:secular_evolution_mmr32}
\end{figure*}

In the 3:2 MMR case, the variations in the resonant angles are more symmetric than in the 2:1 MMR case. For the 2:1 MMR, $\phi_2$ is more vulnerable to perturbations from the giant planets, and in most systems $\Delta\phi_2$ reaches $\pi$ (circulation), while $\phi_1$ remains only mildly perturbed and still librating. In contrast, in the 3:2 MMR case the magnitudes of $\Delta\phi_1$ and $\Delta\phi_2$ are more comparable. Figure~\ref{fig:secular_evolution_mmr32} shows the same two examples as in Figure~\ref{fig:secular_evolution_mmr21}, but now the inner SEs start in a 3:2 MMR. Their eccentricity evolution closely resembles the behavior seen in the 2:1 MMR case. However, in both examples (panels c and f), $\phi_1$ and $\phi_2$ exhibit nearly identical libration patterns, in contrast to the asymmetric evolution characteristic of the 2:1 MMR case. 

\begin{figure*}
    \centering
    \includegraphics[width=\textwidth]{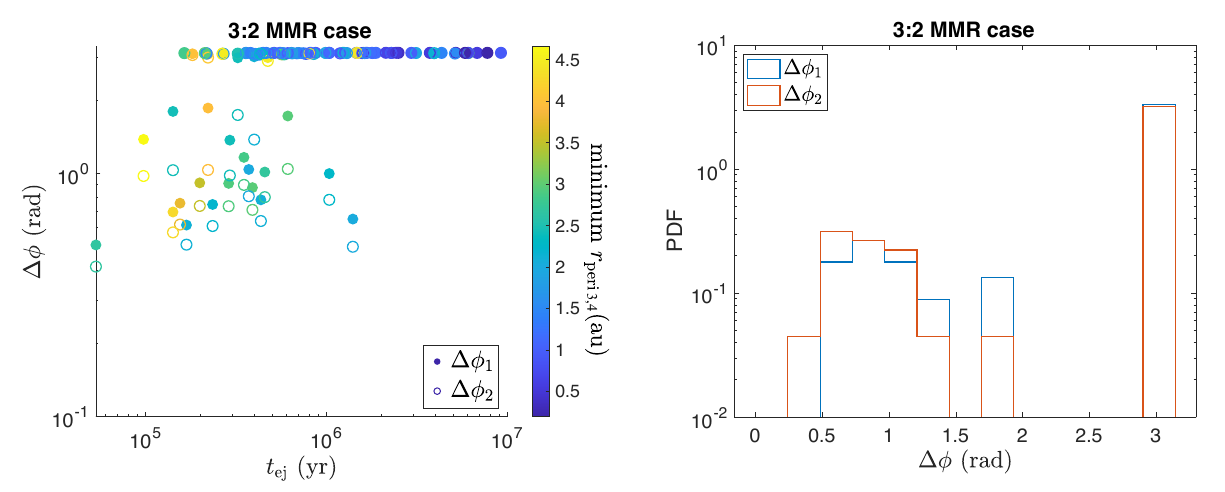}
    \caption{Same as Figure \ref{fig:res_angle_mmr21}, but for the case in which the inner SEs begin in a 3:2 MMR.}
    \label{fig:res_angle_mmr32}
\end{figure*}

Out of 93 systems in which the ejected giant planet’s minimum pericenter distance remains larger than the outer super-Earth’s orbit ($r_{\rm peri,2} > a_2$), 81 systems end with $\phi_1$ circulating and 76 systems end with $\phi_2$ circulating. Figure \ref{fig:res_angle_mmr32} shows the distribution of $\Delta\phi$ as a function of the ejection time $t_{\rm ej}$ and the corresponding PDF. The trend seen in the 2:1 MMR case reappears here: systems with longer $t_{\rm ej}$ tend to reach smaller minimum $r_{\rm peri 3,4}$ and correspondingly larger $\Delta\phi$. Approximately 15\% of the systems show moderate perturbations in both resonant angles ($0 < \Delta\phi_{1,2} < 1.5$), while the remaining $\sim$85\% end with both resonant angles fully circulating. 

\subsubsection{Dependence on the location of the SEs} \label{subsec:a-dependence}

\begin{figure*}
    \centering
    \includegraphics[width=\textwidth]{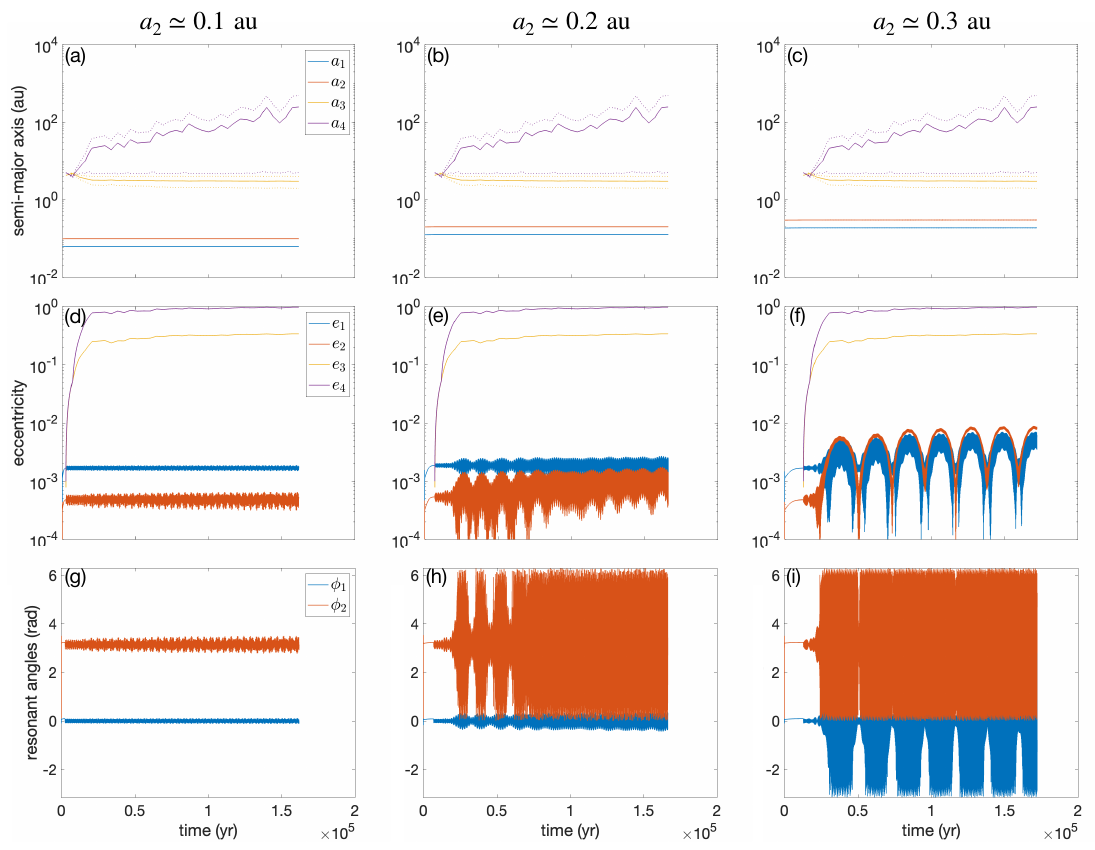}
    \caption{Dependence of the secular impact from the CJs on the location of the SEs (2:1 MMR). Each row shows the time evolution of the semi-major axes (panels a-c), eccentricities (panels d-f), and resonant angles (panels g-i). Each column corresponds to a different value of $a_2$, the location of the outer SE. The color coding is the same as in Figures \ref{fig:secular_evolution_mmr21} and \ref{fig:secular_evolution_mmr32}.}
    \label{fig:a2_dependence_21}
\end{figure*}

\begin{figure*}
    \centering
    \includegraphics[width=\textwidth]{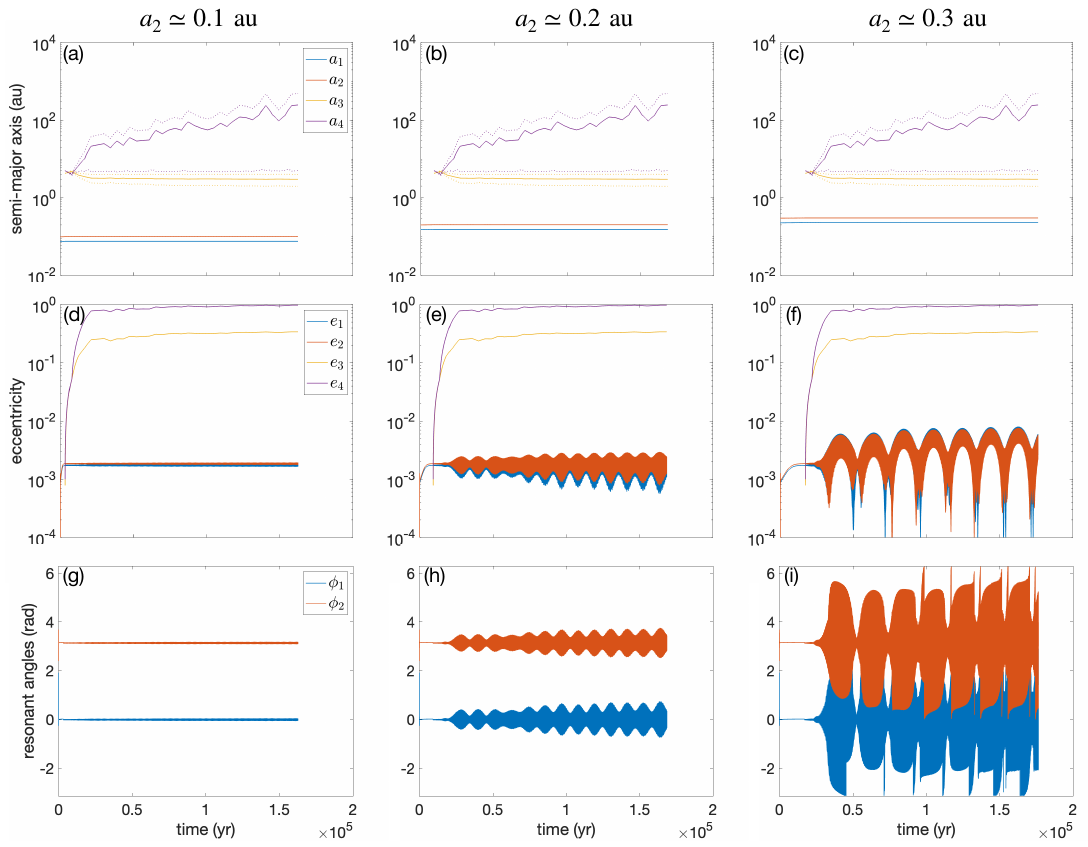}
    \caption{Same as Figure \ref{fig:a2_dependence_21} but for the 3:2 MMR case.}
    \label{fig:a2_dependence_32}
\end{figure*}

Although this work is not intended as a full parameter survey, it is still useful to demonstrate how the scattering influence of CJs on inner super-Earth MMRs depends on the CJs’ orbital distance relative to the inner SE system. To illustrate this dependence, we vary the semi-major axis of the outer super-Earth, $a_2$, from 0.2 au to 0.1 au and 0.3 au, adjusting $a_1$ accordingly to keep the planets in either 2:1 or 3:2 MMR. The time evolution of the giant planets is taken from the “mildly scattering’’ configuration, as shown in the upper panels of Figures \ref{fig:secular_evolution_mmr21} and \ref{fig:secular_evolution_mmr32}.

Figure \ref{fig:a2_dependence_21} shows the evolution of the semi-major axes, eccentricities, and resonant angles of the inner SEs in the 2:1 MMR for the three different values of $a_2$. As the SE pair is placed on progressively wider orbits (i.e., closer to the giant planets), the perturbations they experience become stronger. This leads to more pronounced eccentricity excitation and larger variations in the resonant angles. When $a_2$ increases from 0.2 au to 0.3 au, the previously “moderately’’ perturbed $\phi_1$ (which still librates at 0.2 au) begins to circulate periodically at 0.3 au, driven by the enhanced secular forcing from the giant planets.

The same trend appears in the 3:2 MMR configuration, as shown in Figure \ref{fig:a2_dependence_32}. As $a_2$ increases, the amplitudes of eccentricity excitation and resonant-angle modulation grow, reflecting the stronger secular perturbations from the outer giants. The main difference relative to the 2:1 case is that in the 3:2 MMR, the variations in $\phi_1$ and $\phi_2$ are more symmetric (see Section \ref{subsec:3-2mmr}). Under similar levels of secular forcing, the resonant-angle variations are generally smaller in the 3:2 MMR than in the 2:1 MMR. This is because the 2:1 resonance is intrinsically more vulnerable to secular perturbations: its resonant term is weakened by the indirect potential \citep{Choksi_2023}, making it easier to be disrupted.

\subsection{Validity of the secular approach} \label{subsec:secular_approach}

An important assumption in this work is the use of secular perturbation theory to estimate how outer giant planets undergoing scattering influence the inner SEs. This approach allows us to avoid brute-force long-term integrations of the inner SEs and outer CJs simultaneously, which are time-consuming, given the large difference in their dynamical timescales (see also PL21). This approach is justified for three reasons.

First, the SEs have much smaller masses than the giant planets, so their back-reaction on the giants’ orbits is negligible. Second, direct intrusion of giant planets into the inner system is rare: in fewer than about 10\% of our systems does the minimum heliocentric distance of the giant planet become smaller than 0.1 times its initial semi-major axis (see Section \ref{sec:scatter}). Most systems therefore remain well separated, validating a secular treatment.

Third, the characteristic timescale of close encounters between the giant planets is long compared to the orbital periods of the inner SEs. The encounter timescale can be estimated as $\tau_{\rm enc} \sim r_{34}/v_{34}$, where $r_{34}$ and $v_{34}$ are the separation and relative velocity of the two giant planets. During close encounters, $r_{34}$ is of order the mutual Hill radius $R_{H,34} \sim a_{34}[m_{34}/(3M_*)]^{1/3}$, and the relative velocity is $v_{34} \sim \sqrt{Gm_{34}/r_{34}} \sim n_{34}R_{H,34}$. This gives
\begin{equation}
    \tau_{\rm enc} \sim \frac{R_{H,34}}{n_{34}R_{H,34}} \sim P_{34} \gg P_2, P_1,
    \label{eq:t_enc}
\end{equation}
where $P_{34}$ is the characteristic orbital period of the giant planets and $P_1$, $P_2$ are those of the inner SEs. The clear separation of timescales supports the validity of modeling the inner system’s response using secular perturbation theory.

\section{Summary and Discussion} \label{sec:conclusion}

In this work we have investigated how the scattering history of cold Jupiters (CJs) at a few au influences the inner super-Earth (SE) pairs locked in first-order MMR. Our main findings are as follows:

\begin{itemize}
    \item Critical distance for perturbing inner MMRs with one pericenter passage: 
    The amplitude of resonant-angle variations of the SE pair increases with the perturbing acceleration ratio $\eta$ (see Equation \ref{eq:eta}). In the 2:1 MMR case, $\phi_2$ (see Equation \ref{eq:phi12}) tend to circulate once $\eta \gtrsim 10^{-3}$, whereas $\phi_1$ remains librating until $\eta \gtrsim 0.1$. In the 3:2 MMR case, both resonant angles respond similarly and begin to circulate when $\eta \gtrsim 0.01$. The orbital period ratio remains within 1\% of exact resonance across the explored parameter space ($\eta \lesssim 0.3$).

    \item Frequency of close CJ intrusions from scattering:
    Using REBOUND simulations, we have quantified how often scattering CJs penetrate the inner SE system. In systems with two CJs, fewer than about 10\% of scattering CJs reach within 0.1 times its initial semi-major axis before being ejected. In systems with three CJs, this probability increases to about $20\%$.

    \item Secular impact of scattering CJs on inner SEs in MMR:
    Even without direct close encounters, the time-dependent apsidal forcing from scattering CJs can strongly disturb inner SEs in MMR. Combining secular evolution of SEs with REBOUND-derived forcing from the scattering CJs, we find that, for example, when one CJ ends up on a 3-au orbit and the other is ejected, and the outermost SE lies at $\simeq 0.2$ au, more than 60\% of inner 2:1 systems end with $\phi_1$ circulating and nearly all with $\phi_2$ circulating; for the 3:2 case, roughly 85\% of systems show both angles circulating, reflecting more symmetric resonant responses. The effect strengthens as the separation between the CJ and SE orbits decreases. Thus, even when the ``final'' CJ has a negligible effect on the ``current'' SE orbital architecture, its previous ``violent'' history likely has left a large impact on the orbits of the SEs.

\end{itemize}

These results show that CJ scattering provides an efficient pathway to disrupt inner MMRs, potentially triggering longer-term instabilities and contributing to the observed pile-up of SE period ratios just wide of exact resonances. 

\subsection{Limitations of the coplanar case} \label{subsec:coplanar}

A caveat of our treatment is that, for simplicity, we adopt a coplanar configuration in both the secular analysis and the two–CJ scattering experiments. 
The inner SEs are assumed to be nearly coplanar, consistent with the observed architecture of Kepler multi-planet systems.
Scattering between two CJs is unlikely to excite large mutual inclinations and thus influence the inclinations of the inner SEs.

However, when three or more giant planets participate in the scattering, much larger inclinations can be excited \citep[e.g.,][]{Lu_2024}. In such cases, the highly inclined outer giant(s) may influence the inner architecture \citep{Pu_2018, Pu_2021}, even for SEs in MMR \citep{Rodet_Lai_2021}. 
A full treatment of inclination dynamics is beyond the present scope and will be pursued in future work.

\subsection{Extension to Neptune scattering} \label{subsec:extension}

About 30\% of the SE systems have external CJ companions \citep{Zhu_2018, Bryan_2024}. Thus the effects studied in this paper cannot operate in the majority of the observed SE systems. However, similar process can also operate if the CJs are replaced by Neptune-mass planets.
Compared to CJs, Neptunes are significantly more abundant. While CJ occurrence rate among all solar-type stars is about $10\%$ \citep[e.g.,][]{Wittenmyer_2016, Fulton_2021}, Neptunes occur at a rate of roughly one per three stars \citep{Zang_2025}. 
This suggests that Neptunes may contribute more broadly to shaping inner SE architectures.

Neptune scattering operates on longer dynamical timescales than CJ scattering, owing to their lower masses and typically wider orbits. Ejection generally requires multiple encounters, which allows a Neptune to intrude deeper into the inner system before being removed. This longer interaction phase enables two pathways for perturbing the inner SEs: (i) prolonged residence on high-eccentricity orbits that drive secular excitation, or (ii) close encounters capable of directly shifting orbital periods and disrupting MMR.
Scattering among multiple Neptunes can also generate a wider diversity of final orbital configurations, including substantial inclination excitation \citep{Hadden_2025}. Such high-inclination outcomes may introduce additional secular pathways for disturbing the inner system beyond those accessible in the two–CJ scenario considered in this paper. Thus Neptunes (and their mutual scattering)
may serve as potentially more efficient agents for MMR disruption among inner SEs. In future work, we will extend our analysis to quantify how Neptune scattering modifies inner resonances, evaluate the frequency of disruptive encounters, and determine whether Neptune-driven instability can account for the observed distribution of near-resonant SE pairs.

\begin{acknowledgments}
We thank Prof. Yanqin Wu for valuable discussions. We thank the anonymous reviewer for their valuable comments, which have helped improve the quality of this paper.
This work is supported by National Natural Science Foundation of China (grant No. 12503069).
K. G. acknowledges the support from K. C. Wong Educational Foundation.
Some of the numerical computations were carried out on the general-purpose PC cluster (PCC) at the Center for Computational Astrophysics, National Astronomical Observatory of Japan.
\end{acknowledgments}

%



\appendix

\section{Secular Equations of Motion of inner Super-Earths under secular perturbations} \label{appendix:EoM}

Our model consists of two inner SEs in $(q+1):q$ MMR and two outer CJs undergoing scattering. Under the resonant interaction and secular perturbations from the CJs, the motion of the inner SEs is governed by the following equations:

\begin{align*}
    \dot{n}_1 &= 3qn_1^2\alpha\mu_2(e_1f_1\sin\phi_1-e_2f_2\sin\phi_2)\\
    \dot{e}_1 &= n_1\alpha\mu_2f_1\sin\phi_1-n_1\alpha \mu_2f_4e_2\sin(\varpi_1-\varpi_2) \\
     & \quad -\frac{15}{16}\mu_3n_3^{8/3}n_1^{-5/3}e_3(1-e_3^2)^{-5/2}\sin(\varpi_1-\varpi_3)\\
     & \quad -\frac{15}{16}\mu_4n_4^{8/3}n_1^{-5/3}e_4(1-e_4^2)^{-5/2}\sin(\varpi_1-\varpi_4) \\
     \dot{\varpi}_1 &= -\frac{f_1\alpha n_1\mu_2}{e_1}\cos\phi_1+n_1\alpha \mu_2\left[2f_3-f_4\frac{e_2}{e_1}\cos(\varpi_1-\varpi_2)\right]\\
     & \quad +\mu_3n_1^{-1}n_3^2(1-e_3^2)^{-3/2}\left[\frac{3}{4}-\frac{15}{16}\left(\frac{n_3}{n_1}\right)^{2/3}\frac{e_3}{e_1}(1-e_3^2)^{-1}\cos(\varpi_1-\varpi_3)\right] \\
     & \quad +\mu_4n_1^{-1}n_4^2(1-e_4^2)^{-3/2}\left[\frac{3}{4}-\frac{15}{16}\left(\frac{n_4}{n_1}\right)^{2/3}\frac{e_4}{e_1}(1-e_4^2)^{-1}\cos(\varpi_1-\varpi_4)\right] \\
     \dot{\phi}_1 &= (q+1)n_2 - qn_1 - \dot{\varpi}_1 \\
     \dot{n}_2 &= -3(q+1)n_2^2\mu_1(e_1f_1\sin\phi_1-e_2f_2\sin\phi_2) \\
     \dot{e}_2 &= -n_2\mu_1f_2\sin\phi_2+n_2\mu_1f_4e_1\sin(\varpi_1-\varpi_2) \\
     & \quad -\frac{15}{16}\mu_3n_3^{8/3}n_1^{-5/3}e_3(1-e_3^2)^{-5/2}\sin(\varpi_1-\varpi_3)\\
     & \quad -\frac{15}{16}\mu_4n_4^{8/3}n_1^{-5/3}e_4(1-e_4^2)^{-5/2}\sin(\varpi_1-\varpi_4) \\
     \dot{\varpi}_2 &= \frac{n_2\mu_1f_2}{e_2}\cos\phi_2+n_2\mu_1\left[2f_3-f_4\frac{e_1}{e_2}\cos(\varpi_1-\varpi_2)\right] \\
     & \quad +\mu_3n_1^{-1}n_3^2(1-e_3^2)^{-3/2}\left[\frac{3}{4}-\frac{15}{16}\left(\frac{n_3}{n_1}\right)^{2/3}\frac{e_3}{e_1}(1-e_3^2)^{-1}\cos(\varpi_1-\varpi_3)\right] \\
     & \quad +\mu_4n_1^{-1}n_4^2(1-e_4^2)^{-3/2}\left[\frac{3}{4}-\frac{15}{16}\left(\frac{n_4}{n_1}\right)^{2/3}\frac{e_4}{e_1}(1-e_4^2)^{-1}\cos(\varpi_1-\varpi_4)\right] \\
     \dot{\phi}_2 &= (q+1)n_2 - qn_1 - \dot{\varpi}_2,
\end{align*}
where $n$ is the mean motion (orbital frequency), $\alpha = a_1/a_2$ is the semi-major axis ratio, and the subscripts denote the corresponding quantities for each planet. The coefficients $f_1, f_2, f_3,$ and $f_4$ depend on $\alpha$ and the Laplace coefficients $b_l^m$ \citep{Murray_1999}. We omit their explicit forms and refer the reader to Equations A5–A9 in \citet{Choksi_2023}. For SE pairs in the 2:1 and 3:2 MMRs, the numerical values of these coefficients are listed in Table \ref{tab:coefficients}.

The time-dependent orbital quantities associated with the scattering CJs $n_{3,4}$, $e_{3,4}$, and $\varpi_{3,4}$ are taken from REBOUND simulations as described in Section \ref{subsec:scat_results}.

We numerically integrate the above secular EoM using the \texttt{lsoda} algorithm implemented in \texttt{scipy.integrate.solve\_ivp} \citep{Virtanen_2020}, with relative and absolute tolerances set to $10^{-10}$ and $10^{-14}$, respectively.

\begin{deluxetable}{ccccc}
\tablewidth{0pt}
\tablecaption{Coefficients used in the secular EoM of the SEs: $f_1$ and $f_2$ quantify the resonant forcing, while $f_3$ and $f_4$ characterize the secular coupling between the inner SEs.} \label{tab:coefficients}
\tablehead{
\colhead{} & \colhead{$f_1$} & \colhead{$f_2$} & \colhead{$f_3$} & \colhead{$f_4$}
}
\startdata
2:1 & 1.19 & 0.43 & 0.39 & 0.58\\
3:2 & 2.02 & 2.48 & 1.15 & 2.0 \\
\enddata
\end{deluxetable}

\section{Full N-body simulations} \label{appendix:full_NB}

\begin{figure*}
    \centering
    \includegraphics[width=\textwidth]{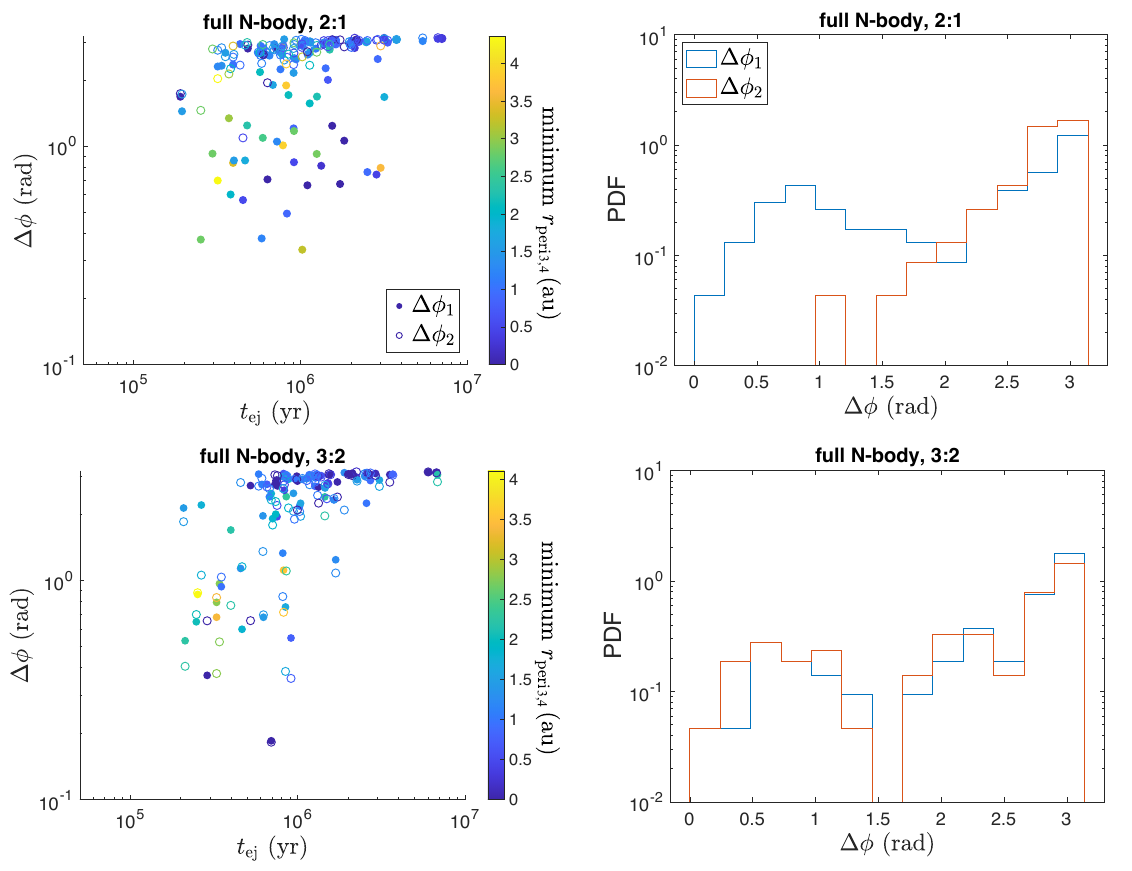}
    \caption{Same as Figures \ref{fig:res_angle_mmr21} (upper) and \ref{fig:res_angle_mmr32} (bottom), but for the results of full N-body simulation. The upper and bottom rows show the cases where the inner SEs are initially in 2:1 and 3:2 MMR, respectively. The initial orbital configuration of the all four planets are the same as in the set up for secular analysis.}
    \label{fig:full_NB}
\end{figure*}

To assess the robustness of our secular analysis, we also perform full N-body simulations that include both the inner SEs and the outer CJs. We adopt the same initial conditions as in the secular models and integrate the orbits of all four planets until one of the giant planets is ejected.

Figure \ref{fig:full_NB} summarizes the outcomes of the full N-body simulations. Systems in which the inner SEs become dynamically unstable are removed from the sample (less than 10\% in each case). The upper and lower rows show the distributions of $\Delta \phi$ versus $t_{\rm ej}$ and the corresponding PDFs of $\Delta \phi$ for the 2:1 and 3:2 MMR cases, respectively. Several similarities and differences relative to the secular results can be identified.

First, the trend that the minimum pericenter distance $r_{\rm peri,2}$ becomes smaller for larger $t_{\rm ej}$—leading to stronger perturbations to the resonant angles (larger $\Delta \phi$)—is consistent with what is shown in the left panels in Figures \ref{fig:res_angle_mmr21} and \ref{fig:res_angle_mmr32}.

Second, the more symmetric change of $\Delta \phi_1$ and $\Delta \phi_2$ in the 3:2 MMR case compared to the 2:1 case is also reproduced in the full N-body simulations, again in agreement with the secular predictions.

The main difference is that the N-body results exhibit a more dispersed distribution of $\Delta \phi$ values than the secular analysis. This is evident from the more continuous PDFs of both $\Delta \phi_1$ and $\Delta \phi_2$ in Figure \ref{fig:full_NB}, relative to the sharper distributions in Figures \ref{fig:res_angle_mmr21} and \ref{fig:res_angle_mmr32}. This broader spread naturally reflects the higher accuracy and greater dynamical richness of the full N-body calculations.

Overall, the full N-body simulations closely track the secular predictions, reinforcing the reliability and applicability of the secular approach. 

\bibliography{references}{}
\bibliographystyle{aasjournalv7}



\end{CJK*}
\end{document}